\documentclass[aps,twocolumn,floatfix,eqsecnum,showpacs,prb]{revtex4}
\usepackage{graphicx}
\usepackage{amsmath}
\usepackage{amssymb}
\usepackage{bm}

\begin{document}

\title{Scattering formula for the topological quantum number of a disordered multi-mode wire}
\author{I. C. Fulga}
\affiliation{Instituut-Lorentz, Universiteit Leiden, P.O. Box 9506, 2300 RA Leiden, The Netherlands}
\author{F. Hassler}
\affiliation{Instituut-Lorentz, Universiteit Leiden, P.O. Box 9506, 2300 RA Leiden, The Netherlands}
\author{A. R. Akhmerov}
\affiliation{Instituut-Lorentz, Universiteit Leiden, P.O. Box 9506, 2300 RA Leiden, The Netherlands}
\author{C. W. J. Beenakker}
\affiliation{Instituut-Lorentz, Universiteit Leiden, P.O. Box 9506, 2300 RA Leiden, The Netherlands}

\date{January, 2011}

\begin{abstract}
The topological quantum number ${\cal Q}$ of a superconducting or chiral insulating wire counts the number of stable bound states at the end points. We determine ${\cal Q}$ from the matrix $r$ of reflection amplitudes from one of the ends, generalizing the known result in the absence of time-reversal and chiral symmetry to all five topologically nontrivial symmetry classes. The formula takes the form of the determinant, Pfaffian, or matrix signature of $r$, depending on whether $r$ is a real matrix, a real antisymmetric matrix, or a Hermitian matrix. We apply this formula to calculate the topological quantum number of $N$ coupled dimerized polymer chains, including the effects of disorder in the hopping constants. The scattering theory relates a topological phase transition to a conductance peak, of quantized height and with a universal (symmetry class independent) line shape. Two peaks which merge are annihilated in the superconducting symmetry classes, while they reinforce each other in the chiral symmetry classes.
\end{abstract}

\maketitle

\section{Introduction}
\label{intro}

The bulk-boundary correspondence in the quantum Hall effect equates the number ${\cal Q}$ of occupied Landau levels in the two-dimensional bulk to the number of propagating states at the edge, which is the quantity measured in electrical conduction.\cite{Hal82,But88} Thouless \textit{et al.} identified ${\cal Q}$ as a topological quantum number,\cite{Tho82} determined by an invariant integral of the Hamiltonian $H(\bm{k})$ over the Brillouin zone. 

One-dimensional wire geometries can also be classified by a topological quantum number, which then counts the number of stable (``topologically protected'') bound states at the end points. Examples exist in chiral insulators (such as a dimerized polyacetylene chain\cite{Jac81}) and in superconductors (such as a chiral \textit{p}-wave wire\cite{Kit01}). In the former case the end states are half-integer charged solitons, in the latter case they are charge-neutral Majorana fermions.

Following the line of thought from the quantum Hall effect, one might ask whether the number ${\cal Q}$ of these end states can be related to a transport property (electrical conduction for the insulators and thermal conduction for the superconductors). The basis for such a relationship would be an alternative formula for ${\cal Q}$, not in terms of $H({\bm k})$,\cite{Kit01,Qi10,Ber10,Gho10,Sch10,Lor10} but in terms of the scattering matrix $S$ of the wire, connected at the two ends to electron reservoirs.

This analysis was recently carried out for the superconducting \textit{p}-wave wire,\cite{Akh10} which represents one of the five symmetry classes with a topologically nontrivial phase in a wire geometry.\cite{Has10,Qi10b} In this paper we extend the scattering theory of the topological quantum number to the other four symmetry classes, including the polyacetylene chain as an application.

The outline is as follows. In the next section we show how to construct a topological invariant ${\cal Q}$ from the reflection matrix $r$ (which is a subblock of $S$). Depending on the presence or absence of particle-hole symmetry, time-reversal symmetry, spin-rotation symmetry, and chiral (or sublattice) symmetry, this relation takes the form of a determinant, Pfaffian, or matrix signature (being the number of negative eigenvalues), see Table \ref{tab:table1}. In Sec.\ \ref{endstatenumber} we demonstrate that this ${\cal Q}$ indeed counts the number of topologically protected end states. The connection to electrical or thermal conduction is made in Sec.\ \ref{superchiral}, where we contrast the effect of disorder on the conductance in the superconducting and chiral insulating symmetry classes. We conclude in Sec.\ \ref{poly} with the application to polyacetylene.

\section{Topological quantum number from reflection matrix}
\label{Qrformulas}

\begin{table*}
\centering
\begin{tabular}{ | l || c | c | c | c | c | c | c | }
\hline
symmetry class & D & DIII & BDI & AIII & CII  \\ \hline
topological phase & $\mathbb{Z}_{2}$ & $\mathbb{Z}_{2}$ & $\mathbb{Z}$ & $\mathbb{Z}$ & $\mathbb{Z}$ \\ \hline\hline
particle-hole symmetry & \multicolumn{3}{|c|}{$S=S^{\ast}$} & $\times$ & $S=\Sigma_{y}S^{\ast}\Sigma_{y}$ \\ \hline
time-reversal symmetry & $\times$ & $S=-S^{\rm T}$ & $S=S^{\rm T}$ & $\times$ & $S=\Sigma_{y}S^{\rm T}\Sigma_{y}$ \\ \hline
spin-rotation symmetry & \multicolumn{2}{|c|}{$\times$} & $\checkmark$ & $\checkmark$ or $\times$ & $\times$ \\ \hline
chiral symmetry &  $\times$ & $S^{2}=-1$ & \multicolumn{3}{|c|}{$S^{2}=1$} \\ \hline \hline
reflection matrix & $r=r^{\ast}$ & $r=r^{\ast}=-r^{\rm T}$ & $r=r^{\ast}=r^{\rm T}$ & $r=r^{\dagger}$ & $r=r^{\dagger}=\Sigma_{y}r^{\rm T}\Sigma_{y}$ \\ \hline
topological quantum number & ${\rm sign}\,{\rm Det}\,r$ & ${\rm sign}\,{\rm Pf}\,ir$ & $\nu(r)$ & $\nu(r)$ & $\frac{1}{2}\nu(r)$ \\[2pt] \hline
\end{tabular}
\caption{Classification of the symmetries of the unitary scattering matrix $S$ at the Fermi level in an $N$-mode wire geometry, and relation between the topological quantum number ${\cal Q}$ and the reflection submatrix $r$. For $\mathbb{Z}_{2}$ topological phases ${\cal Q}$ is given in terms of the sign of the determinant (${\rm Det}$) or Pfaffian (${\rm Pf}$) of $r$. For $\mathbb{Z}$ topological phases the relation is in terms of the number $\nu$ of negative eigenvalues of $r$.}
\label{tab:table1}
\end{table*}

The classification of topological phases is commonly given in terms of the Hamiltonian of a closed system.\cite{Ryu10} For the open systems considered here, the scattering matrix provides a more natural starting point. In an $N$-mode wire the scattering matrix $S$ is a $2N\times 2N$ unitary matrix, relating incoming to outgoing modes. The presence or absence, at the Fermi energy $E_{F}$, of particle-hole symmetry, time-reversal symmetry, spin-rotation symmetry, and chiral (or sublattice) symmetry restricts $S$ to one of ten subspaces of the unitary group ${\cal U}(2N)$. In a one-dimensional wire geometry, five of these Altland-Zirnbauer symmetry classes\cite{Alt97} can be in a topological phase, distinguished by an integer-valued quantum number ${\cal Q}$. 

The symmetries of the scattering matrix in the five topological symmetry classes are summarized in Table \ref{tab:table1}. For each class we have chosen a basis for the incoming and outgoing modes at the Fermi level in which the symmetry relations have a simple form. (In the next section we will be more specific about the choice of basis.) Notice that the chiral symmetry operation is the combination of particle-hole and time-reversal symmetry (if both are present).

Topological phases are characterized by a resonance at the Fermi level, signaling the presence of one or more quasi-bound states at the end-points of the wire with vanishingly small excitation energy. (If the wire is superconducting, these excitations are Majorana fermions.\cite{Kit01}) It is therefore natural to seek a relation between the topological quantum number ${\cal Q}$ and the reflection matrix, which is an $N\times N$ submatrix relating incoming and reflected modes from one end of the wire,
\begin{equation}
S=\begin{pmatrix}
r&t'\\
t&r'
\end{pmatrix}.\label{Sdef}
\end{equation}

The wire has two ends, so there are two reflection matrices $r$ and $r'$. Unitarity ensures that the Hermitian matrix products $rr^{\dagger}$ and $r'r'^{\dagger}$ have the same set of reflection eigenvalues $\tanh^{2}\lambda_{n}\in(0,1)$, numbered by the mode index $n=1,2,\ldots N$. The real number $\lambda_{n}$ is the socalled Lyapunov exponent. The transmission eigenvalues $T_{n}=1-\tanh^{2}\lambda_{n}=1/\cosh^{2}\lambda_{n}$ determine the conductance $G\propto \sum_{n}T_{n}$ of the wire. (Depending on the system, this can be a thermal or an electrical conductance.) The topological phases have an excitation gap, so the $T_{n}$'s are exponentially small in general, except when the gap closes at a transition between two topological phases. A topological phase transition can therefore be identified by a sign change of a Lyapunov exponent.\cite{Bro98,Mot01,Mer02,Gru05}

The Lyapunov exponents are the radial variables of the polar decomposition of the scattering matrix, given by\cite{note1}
\begin{widetext}
\begin{subequations}\label{polar}
\begin{align}
&S=\begin{pmatrix}
O_{1}&0\\
0&O_{2}
\end{pmatrix}\begin{pmatrix}
\tanh\Lambda&(\cosh\Lambda)^{-1}\\
(\cosh\Lambda)^{-1}&-\tanh\Lambda
\end{pmatrix}\begin{pmatrix}
O_{3}&0\\
0&O_{4}
\end{pmatrix},\;\;\text{in class D,}\label{SOLambdaD}\\
&S=\begin{pmatrix}
O_{1}&0\\
0&O_{2}
\end{pmatrix}\begin{pmatrix}
(\tanh\Lambda)\otimes i\sigma_{y}&(\cosh\Lambda)^{-1}\otimes i\sigma_{y}\\
(\cosh\Lambda)^{-1}\otimes i\sigma_{y}&-(\tanh\Lambda)\otimes i\sigma_{y}
\end{pmatrix}\begin{pmatrix}
O_{1}^{\rm T}&0\\
0&O_{2}^{\rm T}
\end{pmatrix},\;\;\text{in class DIII,}\label{SOLambdaDIII}\\
&S=\begin{pmatrix}
O_{1}&0\\
0&O_{2}
\end{pmatrix}\begin{pmatrix}
\tanh\Lambda&(\cosh\Lambda)^{-1}\\
(\cosh\Lambda)^{-1}&-\tanh\Lambda
\end{pmatrix}\begin{pmatrix}
O_{1}^{\rm T}&0\\
0&O_{2}^{\rm T}
\end{pmatrix},\;\;\text{in class BDI,}\label{SOBDI}\\
&S=\begin{pmatrix}
U_{1}&0\\
0&U_{2}
\end{pmatrix}\begin{pmatrix}
\tanh\Lambda&(\cosh\Lambda)^{-1}\\
(\cosh\Lambda)^{-1}&-\tanh\Lambda
\end{pmatrix}\begin{pmatrix}
U_{1}^{\dagger}&0\\
0&U_{2}^{\dagger}
\end{pmatrix},\;\;\text{in class AIII,}\label{SOAIII}\\
&S=\begin{pmatrix}
Q_{1}&0\\
0&Q_{2}
\end{pmatrix}\begin{pmatrix}
(\tanh\Lambda)\otimes\sigma_{0}&(\cosh\Lambda)^{-1}\otimes\sigma_{0}\\
(\cosh\Lambda)^{-1}\otimes\sigma_{0}&-(\tanh\Lambda)\otimes\sigma_{0}
\end{pmatrix}\begin{pmatrix}
Q_{1}^{\dagger}&0\\
0&Q_{2}^{\dagger}
\end{pmatrix},\;\;\text{in class CII,}\label{CII}
\end{align}
\end{subequations}
\end{widetext}
in terms of a real diagonal matrix $\Lambda={\rm diag}\,(\lambda_{1},\lambda_{2},\ldots)$ and complex unitary matrices $U_{p}$ (satisfying $U_{p}^{-1}=U_{p}^{\dagger}$), real orthogonal matrices $O_{p}$ (satisfying $O_{p}^{-1}=O_{p}^{\dagger}=O_{p}^{\rm T}$), and quaternion symplectic matrices $Q_{p}$ (satisfying $Q_{p}^{-1}=Q_{p}^{\dagger}=\Sigma_{y}Q_{p}^{\rm T}\Sigma_{y}$). The matrices $\Sigma_{i}=\sigma_{i}\oplus\sigma_{i}\oplus\cdots\oplus\sigma_{i}$ are block diagonal in terms of $2\times 2$ Pauli matrices $\sigma_{i}$ (with $\sigma_{0}$ the $2\times 2$ unit matrix). There are $N$ distinct $\lambda_{n}$'s in classes D, BDI, and AIII, but only $N/2$ in classes DIII and CII (because of a twofold Kramers degeneracy of the transmission eigenvalues). 

The transmission eigenvalues only determine the Lyapunov exponents up to a sign. To fix the sign, we demand in class D and DIII that ${\rm Det}\,O_{p}=1$, so $O_{p}\in{\rm SO}(N)$. Then the $\lambda_{n}$'s can be ordered uniquely as\cite{Gru05} $|\lambda_{1}|<\lambda_{2}<\lambda_{3}<\cdots$, so there can be at most a single negative Lyapunov exponent. In the other three classes there is no sign ambiguity since $\tanh\lambda_{n}$ is an eigenvalue of the reflection matrix $r$ itself --- which is a Hermitian matrix in classes BDI, AIII, and CII. There is then no constraint on the number of negative Lyapunov exponents.\cite{Bro98}

If we start from an initial state with all $\lambda_{n}$'s positive, then the number ${\cal Q}$ of (distinct) negative Lyapunov exponents in a final state counts the number of topological phase transitions that separate initial and final states. In class D this produces the relation ${\cal Q}={\rm sign}\,{\rm Det}\,r$ from Refs.\ \onlinecite{Mer02,Akh10}, relating topological quantum number and determinant of reflection matrix.

In class DIII the determinant of $r$ is always positive, but we can use the Pfaffian of the antisymmetric reflection matrix to count the number of negative $\lambda_{n}$'s, so we take ${\cal Q}={\rm sign}\,{\rm Pf}\,r$. [In view of the identity ${\rm Pf}\,XYX^{T}=({\rm Det}\,X)({\rm Pf}\,Y)$, one has ${\rm Pf}\,r=({\rm Det}\,O_{1}){\rm Pf}\,(\Lambda\otimes i\sigma_{y})=\prod_{n}\tanh\lambda_{n}$.]

In classes BDI and AIII the matrix signature ${\cal Q}=\nu(r)$ of the Hermitian matrix $r$ gives the number of negative eigenvalues, equal to the number of negative $\lambda_{n}$'s. In class CII we take ${\cal Q}=\frac{1}{2}\nu(r)$ to obtain the number of distinct negative $\lambda_{n}$'s, because each eigenvalue is twofold degenerate.

These topological quantum numbers are defined relative to a particular reference state, chosen to have all positive Lyapunov exponents. We would like to relate ${\cal Q}$ to the number of end states at zero excitation energy, and then chose a reference state such that this relationship takes a simple form. This is worked out in the next section, with the resulting expressions for ${\cal Q}$ given in Table \ref{tab:table1}.

\section{Number of end states from topological quantum number}
\label{endstatenumber}

We consider first the superconducting symmetry classes D and DIII and then the chiral symmetry classes BDI, AIII, and CII. The symmetry class D was treated in detail in Ref.\ \onlinecite{Akh10} and is included here for completeness and for comparison with class DIII.

\subsection{Superconducting symmetry classes}
\label{superclass}

Electron-hole symmetry in a superconductor relates the energy-dependent creation and annihilation operators by $\gamma^{\dagger}(E)=\gamma(-E)$. Since therefore $\gamma^{\dagger}=\gamma$ at $E=0$, an excitation at zero energy is a Majorana fermion, equal to its own antiparticle. The end states in symmetry classes D and DIII are socalled Majorana bound states.\cite{Kit01} In the open systems considered here, where the superconducting wire is connected to semi-infinite normal-metal leads, the end states are actually only quasi-bound states, but they still manifest themselves as a resonance in a conduction experiment.\cite{Law09,Wim11}

The topological quantum number in class D should give the parity of the number ${\cal N}$ of Majorana bound states at one end of the wire: ${\cal N}$ is even (${\cal Q}=1$) in the topologically trivial phase, while ${\cal N}$ is odd (${\cal Q}=-1$) in the topologically nontrivial phase. In class DIII all states are twofold Kramers degenerate so ${\cal N}$ is to be replaced by ${\cal N}/2$.

\begin{figure}[tb]
\centerline{\includegraphics[width=0.6\linewidth]{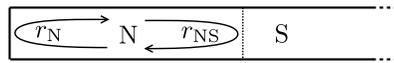}}
\caption{\label{fig_layout}
Superconducting wire (S) connected to a normal-metal lead (N) which is closed at one end. A bound state at the Fermi level can form at the NS interface, characterized by a unit eigenvalue of the product $r_{\rm N}r_{\rm NS}$ of two matrices of reflection amplitudes (indicated schematically by arrows). 
}
\end{figure}

Let us now verify that the determinant and Pfaffian expressions for the topological charge in Table \ref{tab:table1} indeed give this bound state parity. We transform the quasi-bound states into true bound states by terminating the normal-metal lead at some distance far from the normal-superconductor (NS) interface (see Fig.\ \ref{fig_layout}). For the same purpose we assume that the superconducting wire is sufficiently long that transmission of quasiparticles from one end to the other can be neglected. The reflection matrix $r_{\rm NS}$ from the NS interface is then an $N\times N$ unitary matrix. The number of modes $N=2M$ is even, because there is an equal number of electron and hole modes.

The condition for a bound state at the Fermi level is
\begin{equation}
{\rm Det}\,(1-r_{\rm N}r_{\rm NS})=0,\label{boundstaterNrNS}
\end{equation}
where $r_{\rm N}$ is the reflection matrix from the terminated normal-metal lead. In the electron-hole basis the matrix $r_{\rm N}$ has the block-diagonal form
\begin{equation}
r_{\rm N}=\begin{pmatrix}
U_{\rm N}&0\\
0&U_{\rm N}^{\ast}
\end{pmatrix}.\label{rNdef}
\end{equation}
The matrix $U_{\rm N}$ is an $M\times M$ unitary matrix of electron reflection amplitudes. The corresponding matrix for hole reflections is $U_{\rm N}^{\ast}$ because of particle-hole symmetry at the Fermi level.

The reflection matrix from the NS interface has also off-diagonal blocks,
\begin{equation}
r_{\rm NS}=\begin{pmatrix}
r_{ee}&r_{eh}\\
r_{he}&r_{hh}
\end{pmatrix}.\label{rNSdef}
\end{equation}
Particle-hole symmetry relates the complex reflection matrices $r_{he}=r_{eh}^{\ast}$ (from electron to hole and from hole to electron) and $r_{ee}=r_{hh}^{\ast}$ (from electron to electron and from hole to hole). 

\subsubsection{Class D}
\label{subsubD}

A unitary transformation,
\begin{equation}
r=\Omega r_{\rm NS}\Omega^{\dagger},\;\;\Omega=\sqrt{\frac{1}{2}}\begin{pmatrix}
1&1\\
-i&i
\end{pmatrix},\label{rMdef}
\end{equation}
produces a real reflection matrix $r=r^{\ast}$. This is the socalled Majorana basis used for class D in Table \ref{tab:table1}. The determinant is unchanged by the change of basis, ${\rm Det}\,r_{\rm NS}={\rm Det}\,r$.

The condition \eqref{boundstaterNrNS} for a bound state reads, in terms of $r$,
\begin{equation}
{\rm Det}\,(1+O_{\rm N}r)=0,\label{boundstater}
\end{equation}
with $O_{\rm N}=-\Omega r_{\rm N}\Omega^{\dagger}$ an orthogonal matrix. The number ${\cal N}$ of bound states is the number of eigenvalues $-1$ of the $2M\times 2M$ orthogonal matrix $O_{\rm N}r$, while the other $2M-{\cal N}$ eigenvalues are either equal to $+1$ or come in conjugate pairs $e^{\pm i\phi}$. Hence ${\rm Det}\,O_{\rm N}r=(-1)^{\cal N}$ and since ${\rm Det}\,O_{\rm N}=1$ we conclude that ${\rm Det}\,r=(-1)^{\cal N}$, so indeed the determinant of the reflection matrix gives the bound state parity in class D.

\subsubsection{Class DIII}
\label{subsubDIII}

Time-reversal symmetry in class DIII requires 
\begin{equation}
A_{\rm NS}\equiv i\Sigma_{y}r_{\rm NS}=-A_{\rm NS}^{\rm T},\label{rselfdual}
\end{equation}
with $\Sigma_{y}=\sigma_{y}\oplus\sigma_{y}\oplus\cdots\oplus\sigma_{y}$. Instead of Eq.\ \eqref{rMdef} we now define
\begin{equation}
r=\Omega A_{\rm NS}\Omega^{\rm T}.\label{rMselfdual}
\end{equation}
(The matrix $\Sigma_{y}$  acts on the spin degree of freedom, hence it commutes with $\Omega$, which acts on the electron-hole degree of freedom.) In this basis $r=r^{\ast}$ is still real, as required by particle-hole symmetry, while the time-reversal symmetry requirement reads $r=-r^{\rm T}$. This is the basis used for class DIII in Table \ref{tab:table1}. 

The Pfaffians in the two bases are related by ${\rm Pf}\,r=({\rm Det}\,\Omega)({\rm Pf}\,A_{\rm NS})=(-1)^{N/4}\,{\rm Pf}\,A_{\rm NS}$. (Each electron and each hole mode has a twofold Kramers degeneracy, so the total number of modes $N$ is an integer multiple of four.) The relation can be written equivalently as
\begin{equation}
{\rm Pf}\,ir={\rm Pf}\,A_{\rm NS}.\label{Pfaffianrelation}
\end{equation}
This identity is at the origin of the factor $i$ appearing in the class DIII expression for the topological quantum number in Table \ref{tab:table1}.

The condition \eqref{boundstaterNrNS} for a bound state can be rewritten as
\begin{equation}
{\rm Det}\,(A_{\rm N}-r)=\bigl[{\rm Pf}\,(A_{\rm N}-r)]^{2}=0,\label{boundstaterA}
\end{equation}
where $A_{\rm N}\equiv \Omega (i\Sigma_{y}r_{N}^{\dagger})\Omega^{\rm T}$, as well as $r$, are antisymmetric orthogonal matrices. In App.\ \ref{A_Pfaffian} we show that the multiplicity ${\cal N}$ of the number of solutions to Eq.\ \eqref{boundstaterA} satisfies
\begin{equation}
(-1)^{{\cal N}/2}=({\rm Pf}\,A_{\rm N})({\rm Pf}\,r).\label{NPfANPfr} 
\end{equation}
Since, in view of Eq.\ \eqref{rNdef},
\begin{equation}
{\rm Pf}\,A_{N}=({\rm Det}\,\Omega)|{\rm Pf}\,(i\Sigma_{y}U_{\rm N}^{\dagger})|^{2}=(-1)^{N/4},\label{PfAN1}
\end{equation}
we conclude that $(-1)^{N/4}\,{\rm Pf}\,r\equiv {\rm Pf}\,ir$ gives the parity of the number ${\cal N}/2$ of Kramers degenerate bound states. This is the topological quantum number for class DIII listed in Table \ref{tab:table1}.

\subsection{Chiral symmetry classes}
\label{chiralclass}

In the chiral symmetry classes BDI, AIII, and CII we wish to relate the number $\nu(r)$ of negative eigenvalues of the reflection matrix $r$ to the number of quasi-bound states at the end of the wire. As before, we transform these end states into true bound states by terminating the wire and assume that the transmission probability through the wire is negligibly small (so $r$ is unitary). While in the superconducting symmetry classes we could choose a normal metal lead as a unique termination, in the chiral classes there is more arbitrariness in the choice of the unitary reflection matrix $r_{0}$ of the termination.

Since reflection matrices in the chiral classes are Hermitian (see Table \ref{tab:table1}), we can decompose
\begin{equation}
  r_{0} =U_{0}{\cal S}_{n_{0}}
  U_{0}^{\dagger},\;\;{\cal S}_{n_{0}}=
\begin{pmatrix} 
    \openone_{N-n_{0}} & 0 \\
    0       & - \openone_{n_{0}}
   \end{pmatrix},
  \label{r0nu0}
\end{equation}
where $U_{0}$ is an $N\times N$ unitary matrix, $n_{0}=\nu(r_{0})$, and $\openone_{n_{0}}$ is an $n_{0}\times n_{0}$ unit matrix. (Unitarity restricts the eigenvalues to $\pm 1$.) Similarly,
\begin{equation}
  r =U_{1}{\cal S}_{n_{1}}U_{1}^{\dagger},\label{rnu}
\end{equation}
with $\nu(r)=n_{1}$.

Time-reversal symmetry with (without) spin-rotation symmetry restricts the unitary matrices $U_{0}$ and $U_{1}$ to the orthogonal (symplectic) subgroup, but to determine the number of bound states we only need the unitarity.

The condition ${\rm Det}\,( 1- r_0 r) = 0$ for a zero-energy bound state takes the form
\begin{equation}
{\rm Det}\,( 1- {\cal S}_{n_{0}}U{\cal S}_{n_{1}}U^{\dagger}) = 0,\label{calUcalS}
\end{equation}
with $U=U_{0}^{\dagger}U_{1}$. We seek the minimal multiplicity ${\cal N}$ of the solutions of this equation, for arbitrary $U$. (There may be more solutions for a special choice of $U$, but these do not play a role in the characterization of the topological phase.) 

We divide $U$ into four rectangular subblocks,
\begin{equation}
U=\begin{pmatrix}
M_{N-n_{0},N-n_{1}}&M_{N-n_{0},n_{1}}\\
M_{n_{0},N-n_{1}}&M_{n_{0},n_{1}}
\end{pmatrix},\label{blockcalU}
\end{equation}
where $M_{n,m}$ is a matrix of dimensions $n\times m$. Since
\begin{equation}
1- {\cal S}_{n_{0}}U{\cal S}_{n_{1}}U^{\dagger}=2\begin{pmatrix}
0&M_{N-n_{0},n_{1}}\\
M_{n_{0},N-n_{1}}&0
\end{pmatrix}U^{\dagger},\label{calScalUidentity}
\end{equation}
in view of unitarity of $U$,
the bound state equation \eqref{calUcalS} simplifies to
\begin{equation}
{\rm Det}\,\begin{pmatrix}
0&M_{N-n_{0},n_{1}}\\
M_{n_{0},N-n_{1}}&0
\end{pmatrix}=0.\label{simpleboundstaet}
\end{equation}

For any matrix $M_{n,m}$ with $n<m$ there exist at least $m-n$ independent vectors $v$ of rank $m$ such that $M_{n,m}v=0$. Therefore Eq.\ \eqref{simpleboundstaet} has at least $|n_{0}+n_{1}-N|$ independent solutions, hence
\begin{equation}
{\cal N}=|\nu(r)+\nu(r_{0})-N|.\label{calNnuN}
\end{equation}
This is the required relation between the topological quantum number ${\cal Q}=\nu(r)$ (in class BDI, AIII) or ${\cal Q}=\frac{1}{2}\nu(r)$ (in class CII) and the minimal number of bound states ${\cal N}$ at one end of the wire, for arbitrary termination of the wire. In the special case of termination by a reflection matrix $r_{0}=-\openone_{N}\Rightarrow\nu(r_{0})=N$, the relation takes the simple form ${\cal N}={\cal Q}$ (in class BDI, AIII) and ${\cal N}=2{\cal Q}$ (in class CII).

So far we considered one of the two ends of the wire, with reflection matrix $r$. The other end has reflection matrix $r'=-r$ [see Eq.\ \eqref{polar}], so $\nu(r')=N-\nu(r)$. Termination of that end by a reflection matrix $r'_{0}$ produces a minimal number ${\cal N}'$ of bound states given by
\begin{equation}
{\cal N}'=|\nu(r)-\nu(r'_{0})|.\label{calNprime}
\end{equation}

For $r'_{0}=\openone_{N}\Rightarrow\nu(r'_{0})=0$ we have the simple relation ${\cal N}'={\cal Q}$ (in class BDI, AIII) and ${\cal N}'=2{\cal Q}$ (in class CII). The (minimal) number of bound states at the two ends is then the same, but in general it may be different, depending on how the wire is terminated.\cite{Bro02,note2} This arbitrariness in the chiral symmetry classes is again in contrast to the superconducting classes, where Majorana bound states come in pairs at opposite ends of the wire. 

\section{Superconducting versus chiral symmetry classes}
\label{superchiral}

\begin{figure*}[tb]
\centerline{\includegraphics[width=0.6\linewidth]{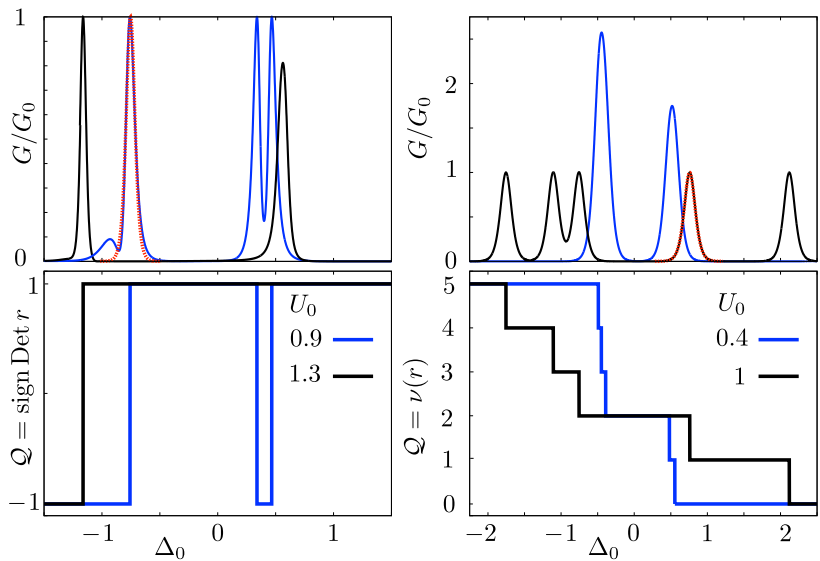}}
\caption{\label{fig_collision}
Conductance $G$ (top panels) and topological quantum number ${\cal Q}$ (bottom panels) in the superconducting class D (left panels) and the chiral class BDI (right panels). The black and blue curves are calculated from the Hamiltonian \eqref{HABDelta}, for a single disorder realization in a wire with $N=5$ modes. The red dotted curve show the universal line shape \eqref{Gpeak} of an isolated conductance peak. Energies $\Delta_{0}$ and $U_{0}$ are measured in units of $\hbar v_{F}/\delta L$ for $\delta L=L/10$.
}
\end{figure*}

As a first application of our general considerations, we contrast the effect of disorder and intermode scattering on topological phase transitions in the superconducting and chiral symmetry classes. We focus on the symmetry classes D and BDI, which in the single-mode case are identical, so that the effect of intermode scattering is most apparent.

In both these classes there is particle-hole symmetry, which implies that we can choose a basis such that the Hermitian Hamiltonian $H$ satisfies
\begin{equation}
H^{\ast}=-H.\label{HclassDsym}
\end{equation}
We assume for simplicity that the $N$ right-moving and left-moving modes all have the same Fermi velocity $v_{F}$. To first order in momentum $p=-i\hbar\partial/\partial x$ the Hamiltonian then takes the form
\begin{align}
 &H =v_{F}p\,\openone_{N}\otimes\sigma_{z} + \Delta_{0}\openone_{N}\otimes\sigma_y\nonumber\\
 &\quad+U_{0}[iA(x)\otimes\sigma_z + iB(x)\otimes\sigma_x + C(x)\otimes\sigma_y],\label{HABDelta}
\end{align}
with $\openone_{N}$ the $N\times N$ unit matrix. The $N\times N$ matrices $A$ and $B$ are real antisymmetric, while $C$ is real symmetric. (For $N=1$ this model Hamiltonian was used in Ref.\ \onlinecite{Akh10}.)

The Hamiltonian \eqref{HABDelta} respects all the symmetries present in class D, but in class BDI the additional chiral symmetry requires
\begin{equation}
\sigma_{x}H\sigma_{x}=-H.\label{classBDIsym}
\end{equation}
This implies that the matrix $B\equiv 0$ in class BDI.

The transfer matrix ${\cal M}$ relates the wave function $\Psi(x)$ at the two ends of the disordered wire (of length $L$): $\Psi(L)={\cal M}\Psi(0)$. At the Fermi level (zero energy) ${\cal M}$ follows upon integration of the wave equation $H\Psi=0$ from $x=0$ to $x=L$,
\begin{align}
&{\cal M}={\cal T}\exp\biggl\{\frac{1}{\hbar v_{F}}\int_{0}^{L}dx\,\biggl(-\Delta_{0}\openone_{N}\otimes\sigma_{x}\nonumber\\
&\quad+U_{0}[A(x)\otimes\sigma_{0}+iB(x)\otimes\sigma_{y}-C(x)\otimes\sigma_{x}]\biggr)\biggr\}.\label{calMdef}
\end{align}
The symbol ${\cal T}$ indicates the ordering of the noncommuting matrices in order of decreasing $x$.

The Pauli matrices in Eq.\ \eqref{calMdef} define a $2\times 2$ block structure for the $2N\times 2N$ transfer matrix. The $N\times N$ reflection matrix $r$ and transmission matrix $t$ follow from this block structure by solving
\begin{equation}
\begin{pmatrix}
t\\0
\end{pmatrix}={\cal M}\begin{pmatrix}
1\\r
\end{pmatrix}.\label{rtM}
\end{equation}
The reflection matrix gives the topological quantum number, ${\cal Q}={\rm sign}\,{\rm Det}\,r$ in class D and ${\cal Q}=\nu(r)$ in class BDI. The transmission matrix gives the conductance
\begin{equation}
G=G_{0}\,{\rm Tr}\,tt^{\dagger}.\label{Gdef} 
\end{equation}
In class D this is a thermal conductance (with $G_{0}=\pi^{2}k_{B}^{2}\tau_{0}/6h$, at temperature $\tau_{0}$), while in class BDI this is an electrical conductance (with $G_{0}=2e^{2}/h$).

We model a disordered wire in class D by taking a Gaussian distribution (zero average, unit variance) of the independent matrix elements of $A(x),B(x),C(x)$, piecewise constant over a series of segments of length $\delta L\ll L$. In class BDI we use the same model with $B\equiv 0$.

In Fig.\ \ref{fig_collision} we plot the conductance and topological quantum number as a function of $\Delta_{0}$ for different values of $U_{0}$, calculated in class D and BDI for a single realization of the disorder. A change in ${\cal Q}$ is accompanied by a peak in $G$, quantized at $G_{0}$ if the topological phase transitions are well separated.\cite{Akh10} The difference between the $\mathbb{Z}_{2}$ superconducting topological phase and the $\mathbb{Z}$ chiral topological phase becomes evident when conductance peaks merge: In the superconducting class D the conductance peaks annihilate, while in the chiral class BDI a maximum of $N$ conductance peaks can reinforce each other.

Also shown in Fig.\ \ref{fig_collision} is that a single isolated conductance peak at $\Delta_{0}=\Delta_{c}$ has the same line shape as a function of $\delta=(\Delta_{0}-\Delta_{c})/\Gamma$,
\begin{equation}
G_{\rm peak}(\delta)=\frac{G_{0}}{\cosh^{2}\delta},\label{Gpeak}
\end{equation}
in both the superconducting and chiral symmetry classes. (The width $\Gamma$ of the peak is not universal.) We have checked that the line shape in the other three symmetry classes also has the same form \eqref{Gpeak}, so this is a general statement. One cannot, therefore, distinguish the $\mathbb{Z}_{2}$ and $\mathbb{Z}$ topological phases by studying a single phase transition. This is a manifestation of the super-universality of Gruzberg, Read, and Vishveshwara.\cite{Gru05}

\section{Application to dimerized polymer chains}
\label{poly}

We conclude with an application in a physical system. Such an application was given for the superconducting symmetry class D in Ref.\ \onlinecite{Akh10}, so here we concentrate on the chiral classes. We consider a dimerized polymer chain such as polyacetylene, with alternating long and short bonds, described by the Su-Schrieffer-Heeger Hamiltonian.\cite{Su79} This is a tight-binding Hamiltonian, which in the continuum limit takes the form of the class BDI Hamiltonian \eqref{HABDelta}.\cite{Jac83} Our goal is to obtain the $\mathbb{Z}$ topological quantum number of $N$ coupled polymer chains from the reflection matrix.

The single-chain electronic Hamiltonian is\cite{Su79,Jac83,Ric82}
\begin{subequations}
\label{HTBpoly}
\begin{align}
&H =-\sum_{n=1}^{N_{L}} t_{n+1,n} (c^\dagger_{n+1}c^{\vphantom{\dagger}}_{n} + c^\dagger_{n}c^{\vphantom{\dagger}}_{n+1}),\label{HTBpolya}\\
&t_{n+1,n}= t_0 - \alpha (u_{n+1}-u_n)\label{HTBpolyb},
\end{align}
\end{subequations}
with $t_0$ and $\alpha$ nearest-neighbor (real) hopping constants and $c_{n}$ the electron annihilation operator at site $n$. (The spin degree of freedom plays no role and is omitted.) Chiral (or sublattice) symmetry means that $H\mapsto -H$ if $c_{n}\mapsto -c_{n}$ on all even-numbered or on all odd-numbered sites. We take $N_{L}$ even, so that the chain contains an equal number of sites on each sublattice. 

Following Jackiw and Semenoff\cite{Jac83} we ignore the atomic dynamics, assuming that the electrons hop in a prescribed atomic displacement field of the dimerized form $u_{n}=(-1)^{n}u_{0}+\delta u_{n}$. Disorder is accounted for by random displacements $\delta u_{n}$, chosen independently on $N$ parallel chains. Nearest neighbors on adjacent chains are coupled by an interchain hopping constant $t_{\rm inter}$, which we take non-fluctuating for simplicity.

\begin{figure}[tb]
\centerline{\includegraphics[width=0.8\linewidth]{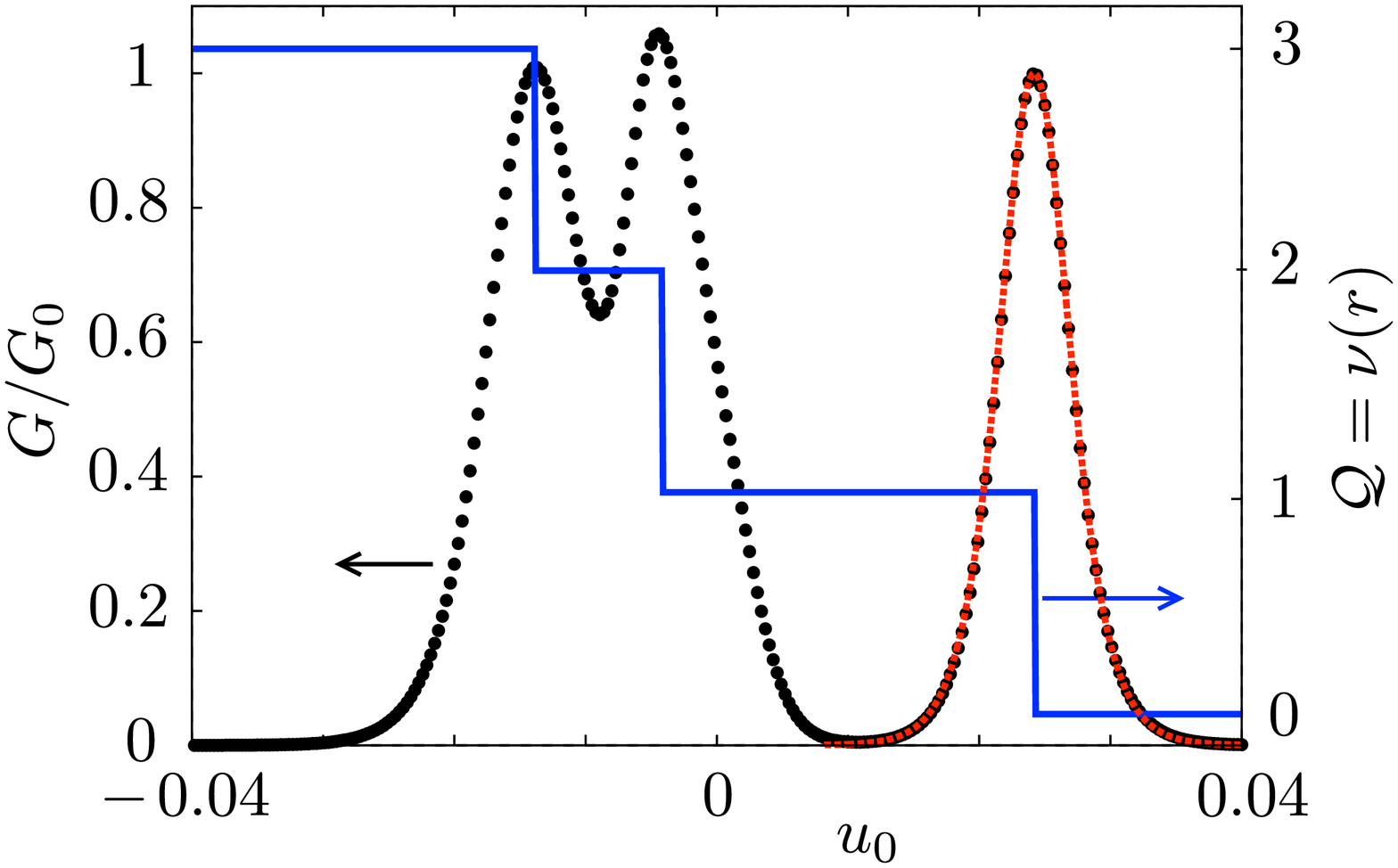}}
\caption{\label{fig_polymer}
Conductance (black dotted line, left axis) and topological quantum number (blue solid line, right axis) of $N=3$ coupled polymer chains (each containing $N_{L}=300$ sites). These curves are calculated from the reflection and transmission matrices, obtained from the Hamiltonian \eqref{HTBpoly} for $t_{0}=1$, $\alpha=1$, and $t_{\rm inter}=0.1$, for a single realization of the random $\delta u_{n}$'s (having a Gaussian distribution with zero average and standard deviation $0.2$). The red dotted curve shows the universal line shape \eqref{Gpeak} of an isolated conductance peak.
}
\end{figure}

The reflection and transmission matrices $r$ and $t$ were computed from the Hamiltonian \eqref{HTBpoly} via the transfer matrix, as outlined in App.\ \ref{app_polyac}. In Fig.\ \ref{fig_polymer} we show the topological quantum number ${\cal Q}$ (equal to the number $\nu(r)$ of negative eigenvalues of the Hermitian reflection matrix $r$), as well as the electrical conductance $G=G_{0}\,{\rm Tr}\,tt^{\dagger}$ (with $G_{0}=2e^{2}/h$). These two quantities are plotted as a function of the dimerization parameter $u_{0}$, to illustrate the topological phase transition, but unlike the excitation gap $\Delta_{0}$ in a superconducting wire this is not an externally controllable parameter.

The case $N=3$ plotted in Fig.\ \ref{fig_polymer} is a $\mathbb{Z}$ topological phase, and each change in the topological quantum number is accompanied by a peak of quantized conductance. The lineshape again has the universal form \eqref{Gpeak}.

\acknowledgments

This research was supported by the Dutch Science Foundation NWO/FOM and by an ERC Advanced Investigator Grant.

\appendix

\section{Calculation of the number of end states in class DIII}
\label{A_Pfaffian}

We wish to prove that the multiplicity ${\cal N}$ of the number of solutions of the bound state equation \eqref{boundstaterA} satisfies Eq.\ \eqref{NPfANPfr}, for arbitrary antisymmetric orthogonal matrices $A_{\rm N}$ and $r$ of dimension $N\times N$, with $N=2M$ and $M$ an even integer. 

We use that any antisymmetric orthogonal matrix can be factorized as $A_{\rm N}=O_{\rm N}i\Sigma_{y}O_{\rm N}^{\rm T}$, $r=O_{\rm NS}i\Sigma_{y}O_{\rm NS}^{\rm T}$, in terms of orthogonal matrices $O_{\rm N}$ and $O_{\rm NS}$. These factorizations relate a Pfaffian to a determinant, ${\rm Pf}\,A_{\rm N}={\rm Det}\,O_{\rm N}$, ${\rm Pf}\,r={\rm Det}\,O_{\rm NS}$.

We seek the multiplicity ${\cal N}$ of the number of solutions of
\begin{equation}
\left[{\rm Pf}\,(A_{\rm N}-r)\right]^{2}=0
\Leftrightarrow\bigl[{\rm Pf}\,(i\Sigma_{y}-Oi\Sigma_{y}O^{\rm T})\bigr]^{2}=0,\label{multiplicity}
\end{equation}
with $O=O_{\rm N}^{\rm T}O_{\rm NS}$ an orthogonal matrix. 

We consider the secular equation for the twofold degenerate eigenvalues $z_{n}$ of the matrix $i\Sigma_{y}Oi\Sigma_{y}O^{\rm T}$,
\begin{align}
0&={\rm Det}\,(z-i\Sigma_{y}Oi\Sigma_{y}O^{\rm T})={\rm Det}\,(zi\Sigma_{y}+Oi\Sigma_{y}O^{\rm T})\nonumber\\
&=\bigl[{\rm Pf}\,(zi\Sigma_{y}+Oi\Sigma_{y}O^{\rm T})\bigr]^{2}=\left[\prod_{n=1}^{M}(z-z_{n})\right]^{2}\Leftrightarrow\nonumber\\
0&={\rm Pf}\,(zi\Sigma_{y}+Oi\Sigma_{y}O^{\rm T})=c\prod_{n=1}^{M}(z-z_{n})=0.\label{seculareq}
\end{align}
The value $c=1$ of the prefactor follows by sending $z$ to infinity. By filling in $z=0$ we find that
\begin{equation}
{\rm Pf}\,(Oi\Sigma_{y}O^{\rm T})={\rm Det}\,O=\prod_{n=1}^{M}z_{n}.\label{seculareq2}
\end{equation}

The ${\cal N}/2$ bound state solutions have $z_{n}=-1$, the remaining $M-{\cal N}/2$ solutions have either $z_{n}=1$ or conjugate pairs $z_{n}=e^{\pm i\phi}$. Hence 
\begin{equation}
\prod_{n=1}^{M}z_{n}=(-1)^{{\cal N}/2}={\rm Det}\,O=({\rm Pf}\,A_{N})({\rm Pf}\,r),\label{QED}
\end{equation}
as we set out to prove.

\section{Calculation of the topological quantum number of a dimerized polymer chain}
\label{app_polyac}

To simplify the notation we outline the calculation of the topological quantum number for the case $N=1$ of a single polymer chain, when the transmission matrix $r$ is a scalar and we may take ${\cal Q}=\frac{1}{2}(1-{\cal Q}')$ with ${\cal Q}'={\rm sign}\,r\in\{-1,1\}$. (The multi-chain case, with ${\cal Q}=\nu(r)\in\{0,1,2,\ldots N\}$, is analogous.) 

From the tight-binding Hamiltonian \eqref{HTBpoly} we directly read off the zero-energy transfer matrix $\tilde{\cal M}$ in the site basis,
\begin{align}
&\left( \begin{array}{c}
         t_{n+1,n}\psi_n \\
	\psi_{n+1}
        \end{array}
 \right) = \tilde{\cal M}_n \left( \begin{array}{c}
        t_{n,n-1}\psi_{n-1} \\
	\psi_{n}
        \end{array}
 \right),\label{Mtna}\\
&\tilde{\cal M}_n = \left( \begin{array}{cc}
         0 & t_{n+1,n}  \\
	 - 1/t_{n+1,n} & 0
        \end{array}
 \right).\label{Mtnb}
\end{align}
The normalization factors in Eq.\ \eqref{Mtna} have been inserted so that the current operator has the site-independent form ${\cal I}=\sigma_{y}$.

To obtain the scattering matrix we need to transform from the site basis to a basis of left-movers and right-movers, in which the current operator equals $\sigma_{z}$ rather than $\sigma_{y}$. This change of basis is realized by the matrix $\Omega$ from Eq.\ \eqref{rMdef},
\begin{equation}
\Omega^{\rm T}\sigma_{y}\Omega^{\ast}=\sigma_{z}.\label{Omegasigma}
\end{equation}

Multiplying the transfer matrices we find for the entire chain (containing an even number of sites $N_{L}$):
\begin{align}
\tilde{\cal M}&=\tilde{\cal M}_{N_{L}}\tilde{\cal M}_{N_{L}-1}\cdots\tilde{\cal M}_{2}\tilde{\cal M}_{1}=\begin{pmatrix}
X&0\\
0&1/X
\end{pmatrix},\label{Mtildepolymer}\\
{\cal M}&=\Omega^{\rm T}\tilde{\cal M}\Omega^{\ast}=\frac{1}{2X}\begin{pmatrix}
X^{2}+1&X^{2}-1\\
X^{2}-1&X^{2}+1
\end{pmatrix},\label{Mpolymer}
\end{align}
with the definition
\begin{equation}
X=  (-1)^{N_{L}/2}\, \prod_{n=1}^{N_{L}/2} \frac{t_{2n+1,2n}}{t_{2n,2n-1}} .\label{Xdef}
\end{equation}

We obtain the reflection and transmission amplitudes from ${\cal M}$ with the help of Eq.\ \eqref{rtM}. The result is
\begin{equation}
r=\frac{1-X^{2}}{1+X^{2}},\;\;t=\frac{2X}{1+X^{2}},\label{rtX}
\end{equation}
so the topological quantum number is given by
\begin{align}
{\cal Q}'&={\rm sign}\,(1-X^{2})\nonumber\\
&={\rm sign}\,\left( \prod_{n=1}^{N_{L}/2} t^2_{2n,2n-1} - \prod_{n=1}^{N_{L}/2} t^2_{2n+1,2n}\right).\label{Qpolymer}
\end{align}

If all hopping constants are close to $t_{0}>0$ we may simplify this expression to
\begin{equation}
{\cal Q}'={\rm sign}\,\left(\sum_{n=1}^{N_{L}/2} \left[t_{2n,2n-1} - t_{2n+1,2n}\right]\right).\label{Qsimple}
\end{equation}
In the absence of disorder, when $t_{2n,2n-1}=t_{0}-2\alpha u_{0}$, $t_{2n+1,2n}=t_{0}+2\alpha u_{0}$, this reduces further to ${\cal Q}'=-\,{\rm sign}\,\alpha u_{0}$, so we recover the original criterion that the dimerized polymer chain has bound states at the ends if the weaker bond is at the end.\cite{Su79}


\begin{thebibliography}{99}
\bibitem{Hal82} B. I. Halperin, Phys. Rev. B \textbf{25}, 2185 (1982).
\bibitem{But88} M. B\"{u}ttiker, Phys. Rev. B \textbf{38}, 9375 (1988).
\bibitem{Tho82} D. J. Thouless, M. Kohmoto, M. P. Nightingale and M. den Nijs, Phys. Rev. Lett. \textbf{49}, 405 (1982).
\bibitem{Jac81} R. Jackiw and J. R. Schrieffer, Nucl. Phys. B \textbf{190}, 253 (1981).
\bibitem{Kit01} A. Yu. Kitaev, Phys. Usp. \textbf{44} (suppl.), 131 (2001).
\bibitem{Qi10} X.-L. Qi, T. L. Hughes, and S.-C. Zhang, Phys. Rev. B \textbf{81}, 134508 (2010).
\bibitem{Ber10} B. B\'{e}ri, Phys. Rev. B \textbf{81}, 134515 (2010).
\bibitem{Gho10} P. Ghosh, J. D. Sau, S. Tewari, and S. Das Sarma, Phys. Rev. B \textbf{82}, 184525 (2010).
\bibitem{Sch10} A. P. Schnyder and S. Ryu, arXiv:1011.1438.
\bibitem{Lor10} T. A. Loring and M. B. Hastings, arXiv:1005.4883; M. B. Hastings and T. A. Loring, arXiv:1012.1019.
\bibitem{Akh10} A. R. Akhmerov, J. P. Dahlhaus, F. Hassler, M. Wimmer, and C. W. J. Beenakker, Phys. Rev. Lett. \textbf{106} (Jan. 20, 2011). [arXiv:1009.5542]
\bibitem{Has10} M. Z. Hasan and C. L. Kane, Rev. Mod. Phys. \textbf{82}, 3045 (2010).
\bibitem{Qi10b} X.-L. Qi and S.-C. Zhang, arXiv:1008.2026.
\bibitem{Ryu10} S. Ryu, A. P. Schnyder, A. Furusaki, and A. W. W. Ludwig, New J. Phys. \textbf{12}, 065010 (2010).
\bibitem{Alt97} A. Altland and M. R. Zirnbauer, Phys. Rev. B \textbf{55}, 1142 (1997).
\bibitem{Bro98} P. W. Brouwer, C. Mudry, B. D. Simons, and A. Altland, Phys. Rev. Lett. \textbf{81}, 862 (1998); P. W. Brouwer, A. Furusaki, and C. Mudry, Phys. Rev. B \textbf{67}, 014530 (2003).
\bibitem{Mot01} O. Motrunich, K. Damle, and D. A. Huse, Phys. Rev. B \textbf{63}, 224204 (2001).
\bibitem{Mer02} F. Merz and J. T. Chalker, Phys. Rev. B \textbf{65}, 054425 (2002).
\bibitem{Gru05} I. A. Gruzberg, N. Read, and S. Vishveshwara, Phys. Rev. B \textbf{71}, 245124 (2005).
\bibitem{note1} In Eq.\ \eqref{polar} we assumed ${\rm Tr}\,S=0$ for the three chiral symmetry classes (BDI, AIII, CII), which means that there is no imbalance between the number of degrees of freedom of opposite chirality (equal number of sites on each sublattice in the case of a bipartite lattice). More generally, $S$ has $N+p$ eigenvalues equal to $+1$ and $N-p$ eigenvalues equal to $-1$, so ${\rm Tr}\,S=2p$ (with $p\in\{-N,\ldots,-1,0,1,\ldots N\}$). The reflection matrices $r$ and $r'$ then each have $|p|$ fully reflected modes, and the polar decomposition \eqref{polar} applies to the remaining $N-|p|$ modes.
\bibitem{Law09} K. T. Law, P. A. Lee, and T. K. Ng, Phys. Rev. Lett. \textbf{103}, 237001 (2009).
\bibitem{Wim11} M. Wimmer, A. R. Akhmerov, J. P. Dahlhaus, and C. W. J. Beenakker, arXiv:1101.5795.
\bibitem{Bro02} P. W. Brouwer, E. Racine, A. Furusaki, Y. Hatsugai, Y. Morita, and C. Mudry, Phys. Rev. B \textbf{66}, 014204 (2002).
\bibitem{note2} The equality $\nu(r)+\nu(r')=N$ holds for ${\rm Tr}\,S=0$, so for balanced chiralities. More generally, for ${\rm Tr}\,S=2p$ one has $\nu(r)+\nu(r')=N-p$. If the wire is terminated by reflection matrices $r_{0}=-\openone_{N}$, $r'_{0}=\openone_{N}$, the number of end states at the two ends differs by $p$.
\bibitem{Su79} W. P. Su, J. R. Schrieffer, and A. J. Heeger, Phys. Rev. Lett. \textbf{42}, 1698 (1979);
Phys. Rev. B \textbf{22}, 2099 (1980).
\bibitem{Jac83} R. Jackiw and G. Semenoff, Phys. Rev. Lett. \textbf{50}, 439 (1983).
\bibitem{Ric82} M. Rice and G. Mele, Phys. Rev. Lett. \textbf{49}, 1455 (1982).
\end{thebibliography}
\end{document}